\documentclass[
reprint,
showpacs,preprintnumbers,
bibnotes,
 amsmath,amssymb,
 aps,
]{revtex4-1}
\usepackage{multirow}
\usepackage{enumerate}
\usepackage{color}
\usepackage{graphicx}
\usepackage{dcolumn}
\usepackage{bm}
\usepackage{subfigure}

\begin{document}
\newcommand{\red}[1]{\textcolor{red}{#1}}
\newcommand{\green}[1]{\textcolor{green}{#1}}
\newcommand{\blue}[1]{\textcolor{blue}{#1}}
\newcommand{\cyan}[1]{\textcolor{cyan}{#1}}
\newcommand{\purple}[1]{\textcolor{purple}{#1}}
\newcommand{\yellowbox}[1]{\colorbox{yellow}{#1}}
\newcommand{\purplebox}[1]{\colorbox{purple}{#1}}
\newcommand{\yellow}[1]{\textcolor{yellow!70!red}{#1}}
\title{Scalar clouds and the superradiant instability regime of  Kerr-Newman black hole}
\author{Yang Huang}
\email{saisehuang@163.com}
\author{Dao-Jun Liu}%
 \email{djliu@shnu.edu.cn}
 \affiliation{Center for Astrophysics, Shanghai Normal University, 100 Guilin Road, Shanghai 200234, China
}%
\date{\today}

\begin{abstract}
In this paper, we study a physical system that is composed of massive charged scalar field linearly coupled to a charged rotating Kerr-Newman black hole. Given the parameters of  black hole and a specific set of "quantum" numbers, the parameter space of the scalar field, which is a plane spanned by its mass and charge,  is divided  into five partitions by three simple constraint lines and the existence line of scalar clouds.  The physical properties of the system in these partitions are presented. It is found that superradiant instability may be possibly caused only in two of the partitions.  In particular, it is shown that  both the mass and charge of the scalar clouds are bounded in a limited region.   Our results may be used to rapidly judge the possible occurrence  of superradiant instability and  the existence of scalar clouds around a given black hole.
   
\end{abstract}

\pacs{04.30.Nk, 04.70.Bw}
\maketitle

\section{Introduction}
It is known that, contrary to static Schwarzschild black holes,   rotating Kerr  black holes can support linearized stationary scalar configurations in their exterior regions. 
The existence of these stationary configurations is intimately related to superradiance and quasibound states of the scalar field \cite{PhysRevD.86.104026,Hod2013,PhysRevLett.112.221101}.

Generally speaking, the interaction between the central black hole and the massive scalar field will prevent low frequency modes with
\begin{equation}\label{eq:omega_mu}
	0<\omega<\mu
\end{equation}
from escaping to spatial infinity, where $\omega$ and $\mu$ are respectively the  frequency and mass of the scalar field.
For the modes that trapped by the central black hole, if they further satisfy the superradiant condition, that is, 
\begin{equation}\label{superradiant condition}
	0<\omega<\omega_c,
\end{equation}
where $\omega_c$ is the critical frequency  for superradiant scattering,  they are growing in time
and can extract energy and angular momentum from the black hole, thus   trigger the superradiant instability of the back hole. 
However, for those modes with $\omega>\omega_c$, they are decaying in time. 
Precisely at the threshold $\omega=\omega_c$, the imaginary part of the frequency vanishes  and these trapped modes are in equilibrium with the black hole. These scalar modes that do not grow or decay in time are dubbed scalar clouds \cite{PhysRevD.86.104026,Hod2013}.
The study of scalar clouds is not only of interest from the viewpoint of black hole theory, but can be also considered as a step to investigate hairy black holes \cite{PhysRevLett.112.221101}.
In a series of interesting work by Hod \cite{Hod2016181,0264-9381-32-13-134002,Hod2012320}, they find that there is an upper bound for the scalar mass to form a potential well outside a Kerr black hole and thus the mass of stationary scalar clouds in a Kerr space-time are bounded in a limited regime.

Recently, it has been shown that massive charged scalar clouds can also exist in the exterior space-time region of a Kerr-Newman black hole \cite{PhysRevD.90.104024,PhysRevD.90.024051}. Hod has derived an analytical formula which determines the discrete spectrum of scalar field masses, corresponding to the  clouds of the massive charged  scalar fields  in the background of a near-extremal Kerr-Newman black hole \cite{PhysRevD.90.024051}. Benone et al make  numerically  a scan in the parameter space of Kerr-Newman black holes to find the location of the existence lines of clouds with different quantum numbers \cite{PhysRevD.90.104024}.

In this paper, we would like to study the superradiant instability and the existence of scalar clouds in  Kerr-Newman space-time from a  different perspective.  
For a given black hole and a specific set of integer "quantum" numbers, the clouds are only possible  along a one-dimensional subset of the two-dimensional parameters space of massive charged scalar field, called an existence line.  We will keep a firm grasp on the existence line and analyze the parameter space of the scalar field. 

The paper is organized as follows.
In Sec. \ref{sec:ReviewKNBH},  the separation of variables procedure for solving the scalar wave equation in the Kerr-Newman background and the boundary conditions to be imposed in order to obtain bound state solutions are reviewed.
In Sec.\ref{sec:RegimeSRI}, using three simple constraint lines and the existence line of scalar clouds which is obtained by numerical integration, we divide the parameter space of the scalar field, a plane spanned by its mass and charge, into five partitions and present  the physical properties of the scalar-field-black-hole system in these regimes.  Finally, we summary the important results and make some remarks  in the conclusion section.   
Throughout the paper, we use natural units in which $G=c=\hbar=1$. 

\section{Kerr-Newman black holes and charged massive scalar field}\label{sec:ReviewKNBH}
We shall consider a physical system that consists of a massive, charged test scalar field minimally coupled to  a rotating charged black hole. The background space-time is described by the Kerr-Newman line element, which, in standard Boyer-Lindquist coordinates, is given by 
\begin{equation}\label{Kerr Newman metric}
	\begin{aligned}
	ds^2=&-\frac{\Delta}{\rho^2}\left(dt-a\sin^2\theta d\phi\right)^2+\frac{\rho^2}{\Delta}dr^2\\&+\rho^2d\theta^2+\frac{\sin^2\theta}{\rho^2}\left[\left(r^2+a^2\right)d\phi-adt\right]^2,
	\end{aligned}
\end{equation}
where
\begin{equation}\label{rho^2 and Delta function}
	\rho^2\equiv r^2+a^2\sin^2\theta,\;\Delta\equiv r^2-2Mr+a^2+Q^2.
\end{equation}
Here, $M$, $Q$ and $a$ are,  the mass, charge and angular momentum per unit mass of the Kerr-Newman black hole, respectively. Without loss of generality, we only consider the cases in which $Q$ and $a$ are positive numbers. 
The background electromagnetic potential reads
\begin{equation}
	A_\mu=\left(-\frac{Qr}{\rho^2},0,0,\frac{aQr\sin^2\theta}{\rho^2}\right).
\end{equation}
There are two horizons of the black hole, which are located at the zeros of $\Delta$,
\begin{equation}\label{The black hole horizons}
	r_\pm=M\pm\sqrt{M^2-a^2-Q^2}.
\end{equation}

The dynamics of a charged massive scalar field $\Psi$ in the Kerr-Newman spacetime is described by the Klein-Gordon wave equation
\begin{equation}
	\left(\nabla^\alpha-iqA^\alpha\right)\left(\nabla_\alpha-iqA_\alpha\right)\Psi=\mu^2\Psi,
\end{equation}
where $\nabla^{\alpha}$ denotes the covariant derivative in the Kerr-Newman geometry, and $\mu$ and $q$ are the mass and charge of the scalar field, respectively. The above equation can be separated, if we decompose the scalar field as
\begin{equation}
	\Psi(t,r,\theta,\phi)=\sum_{l,m}R_{lm}(r)S_{lm}(\theta)e^{im\phi}e^{-i\omega t},
\end{equation}
where $\omega$ is the conserved frequency of the wave field, and $l$ and $m$ are the spheroidal harmonic index and the azimuthal harmonic index of the mode, respectively.
The angular function $S_{lm}(\theta)$ are the standard spheroidal harmonics which are ruled by
\begin{equation}
	\begin{aligned}
	\frac{1}{\sin\theta}&\frac{d}{d\theta}\left(\sin\theta\frac{dS_{lm}}{d\theta}\right)\\&+\left[K_{lm}+\left(\mu^2-\omega^2\right)a^2\sin^2\theta-\frac{m^2}{\sin^2\theta}\right]S_{lm}=0,
	\end{aligned}
\end{equation}
where $K_{lm}$ are separation constants which can be expressed as
\begin{equation}
	K_{lm}=\sum_{k=0}^{\infty}c_k\left[a^2\left(\mu^2-\omega^2\right)\right]^k,
\end{equation}
where $c_0=l(l+1)$ and other coefficients $c_k$ may be found in Ref.\cite{Olver:2010:NHMF,NIST:DLMF}.
The radial function $R_{lm}$ obey the radial equation
\begin{equation}\label{The radial equation}
	\Delta\frac{d}{dr}\left(\Delta\frac{dR_{lm}}{dr}\right)+UR_{lm}=0,
\end{equation}
where
\begin{equation}
	\begin{aligned}
	U=&\left[\omega(a^2+r^2)-am-qQr\right]^2\\&+\Delta\left[2am\omega-K_{lm}-\mu^2(r^2+a^2)\right].
	\end{aligned}
\end{equation}

What we are interested is the bound state solutions, so an exponentially decaying behavior towards spatial infinity is required. In addition, any state in a black hole background should have a purely ingoing boundary condition at the horizon
(in a frame co-rotating with the horizon). Hence, the physically accepted boundary conditions read
\begin{equation}\label{eqn:k}
	R_{lm}(r)\sim \left\{\
	\begin{aligned}
	&e^{-i(\omega-\omega_c)r_*},\;\mathrm{for}\;\;r\rightarrow r_+,\\&\frac{e^{-\sqrt{\mu^2-\omega^2}r}}{r},\;\;\mathrm{for}\;\;r\rightarrow\infty.
	\end{aligned}
	\right.
\end{equation}
Here,  the tortoise coordinate $r_*$ and the critical frequency $\omega_c$ are, respectively, defined by
\begin{equation}
	\frac{dr_*}{dr}=\frac{r^2+a^2}{\Delta},
\end{equation}
and 
\begin{equation}
\omega_c=m\Omega_H+q\Phi_H,
\end{equation}
where $\Omega_H=\frac{a}{r_+^2+a^2}$ is the horizon angular velocity and $\Phi_H=\frac{Qr_+}{r_+^2+a^2}$ is the horizon  electric potential.

Obviously, for a given scalar field's mass $\mu$ and charge parameter $q$, the boundary conditions (\ref{eqn:k}) single out a discrete set of resonance frequencies indexed by node numbers $n=0,1,2,\cdots$.

\section{ Partitioning the parameter space of scalar field }
\label{sec:RegimeSRI}

\subsection{Three constraint lines in the parameter space}

We define a new radial function \cite{Hod2012320} by
\begin{equation}
	\psi_{lm}\equiv\Delta^{1/2}R_{lm}.
\end{equation}
Then,  Eq.(\ref{The radial equation}) can be rewritten in the form of a Schr\"{o}dinger-like wave equation
\begin{equation}
	\frac{d^2\psi_{lm}}{dr^2}+\left(\omega^2-V\right)\psi_{lm}=0,
\end{equation}
where
\begin{equation}
	\omega^2-V=\frac{U+M^2-a^2-Q^2}{\Delta^2}.
\end{equation}
The asymptotic behavior of the effective potential is given by
\begin{equation}\label{The asymptotic behavior of the effective potential}
	V(r)=\mu^2-\frac{2\left(2M\omega^2-qQ\omega-M\mu^2\right)}{r}+\mathcal{O}\left(\frac{1}{r^2}\right).
\end{equation}
As discussed above, the instability of the Kerr-Newman space-time to massive scalar perturbations is a consequence of massive modes which are trapped inside the effective potential well outside the black hole.
For the effective potential to have a trapping well, its asymptotic derivative must be positive, i.e., $\frac{\mathrm{d} V}{\mathrm{d} r}\rightarrow0^+$ as $r\rightarrow\infty$ \cite{Hod2012320}. Thus, from Eq.(\ref{The asymptotic behavior of the effective potential}), it is needed that 
\begin{equation}\label{constraints for omega}
	\omega^2-\frac{qQ}{2M}\omega-\frac{\mu^2}{2}>0.
\end{equation}
Then, the solutions of the above inequality read
\begin{equation}\label{eq:f1}
	\omega>f(\mu,q)\equiv\frac{qQ}{4M}+\sqrt{\frac{\mu^2}{2}+\frac{q^2Q^2}{16M^2}},
\end{equation}
or 
\begin{equation}\label{eq:f2}
	\omega<\frac{qQ}{4M}-\sqrt{\frac{\mu^2}{2}+\frac{q^2Q^2}{16M^2}}.
\end{equation}
Obviously,  solution (\ref{eq:f2}) should be omitted, due to the fact that the right-hand side of it is always negative.
 
Inspired from the inequations (\ref{eq:omega_mu}), (\ref{superradiant condition}) and (\ref{eq:f1}), for a given black hole and a specific set of quantum numbers $(n,l,m)$,  it would be useful to consider the following three constraint lines in the parameter space of the scalar field ($(\mu,q)$-plane):
\paragraph{$f(\mu,q)=\omega_c.$}
This line is actually  a branch of hyperbolic curve in $(\mu,q)$-plane
\begin{equation}\label{eq:fomgC}
qQ=\frac{AB+\sqrt{A^2+8(B^2-1)M^2\mu^2}}{B^2-1},
\end{equation}
where two dimensionless quantities $A$ and $B$ are respectively defined by
\begin{equation}\label{A and B}
A=4m\,M\Omega_H,\;\;B=1-\frac{4Mr_+}{r_+^2+a^2}.
\end{equation}
\paragraph{$f(\mu,q)=\mu$.}
The line is a simple straight line 
\begin{equation}\label{eq:fmu}
	q Q=M\mu.
\end{equation}
Note that this line has nothing to do with the angular momentum of the black hole and the quantum numbers $(n,l,m)$. Interestingly, the line is just the bound of the necessary  condition obtained in Ref.\cite{Furuhashi01122004} in the regime of $qQ\ll1$ and $M\mu\ll1$, for confining the wave in the potential well. Obviously, this line should not be discussed only in the limit of  $qQ\ll1$ and $M\mu\ll1$.
\paragraph{$\mu=\omega_c$.}
  It is easy to find that this line is also a straight line
 \begin{equation}
  q Q=\frac{r_+^2+a^2}{r_+}\mu- m \frac{a}{r_+}.
 \end{equation}
 
It is worth noting that the above three lines intersect at one point
\begin{equation}\label{P1}
P_1:\;(\mu,q)=(\mu_1,q_1)=\left(\frac{m a}{Mr_+-Q^2},\frac{m M a}{Q(Mr_+-Q^2)}\right).
\end{equation}
Besides, curve \textit{a} and line \textit{c} also intersect at another point
\begin{equation}\label{P2}
P_2:\;\; (\mu,q)=(\mu_2,q_2)=\left(0,- \frac{m a}{Q r_+}\right).
\end{equation}

\subsection{Existence line of scalar clouds}
\begin{figure}
	\centering
	\includegraphics[width=\linewidth]{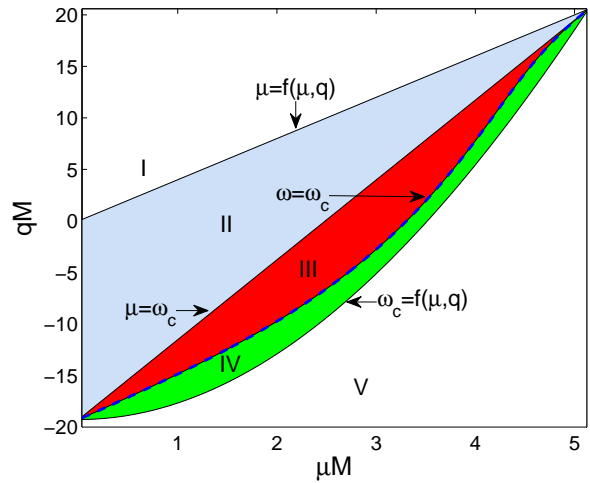}
	\caption{ The parameter space (half of $(\mu,q)$-plane) of scalar field in the background of Kerr-Newman space-time, which is partitioned by three constraint lines \textit{a}, \textit{b}, \textit{c} and the existence line for scalar clouds $\omega=\omega_c$.   Here, the other parameters  are fixed as $n=0$, $l=m=5$, $Q=0.25M$ and $a=0.968235M$.}
	\label{fig:paraplotorder}
\end{figure}
In Fig.\ref{fig:paraplotorder}, the above three lines are plotted in half of  $(\mu,q)$-plane ($\mu\geq 0$),  given a set of  parameters  $(a, Q, l, m)$.
In addition, another curve 
 \begin{equation}\label{omgomgC}
 	\omega=\omega_c,
 \end{equation} 
 which corresponds to the existence line for scalar clouds,  is also plotted.  Different from the three constraint lines, the existence line for scalar clouds can only be obtained by solving  the radial equation (\ref{The radial equation}) with boundary condition (\ref{eqn:k}) as a deformed eigenvalue problem. To this end, we use a numerical method similar to the one used in  Ref.\cite{Furuhashi01122004,PhysRevD.90.104024} , but with a modified strategy, which can be summarized as: 1) first substituting  Eq.(\ref{omgomgC}) into the radial equation (\ref{The radial equation}) and inputing the value of 
 black hole parameters and quantum numbers (to be specific, fixing the value of $(a, Q, l, m)$ and taking the black hole's mass $M$ as a normalization scale); 2) for a given value of $\mu$, integrating the radial equation and obtaining a value of $q$ by a one-parameter shooting procedure; 3) then changing the value of $\mu$ and repeatng step 2).
 After such a process,  $q$ as a function of $\mu$ should be obtained in the end
 \begin{equation}
 q=q( a, Q, n, l, m;\mu).
 \end{equation}
 Note that we are mainly interested in solutions of radial function with $n=0$ in this work. In other words, we perform a nodeless shooting in the practical computation. To gain a solution with nodes, the procedure is similar.  
 
\begin{figure*}    
 	\subfigure { \label{fig:1a}     
 		\includegraphics[width=0.45\textwidth]{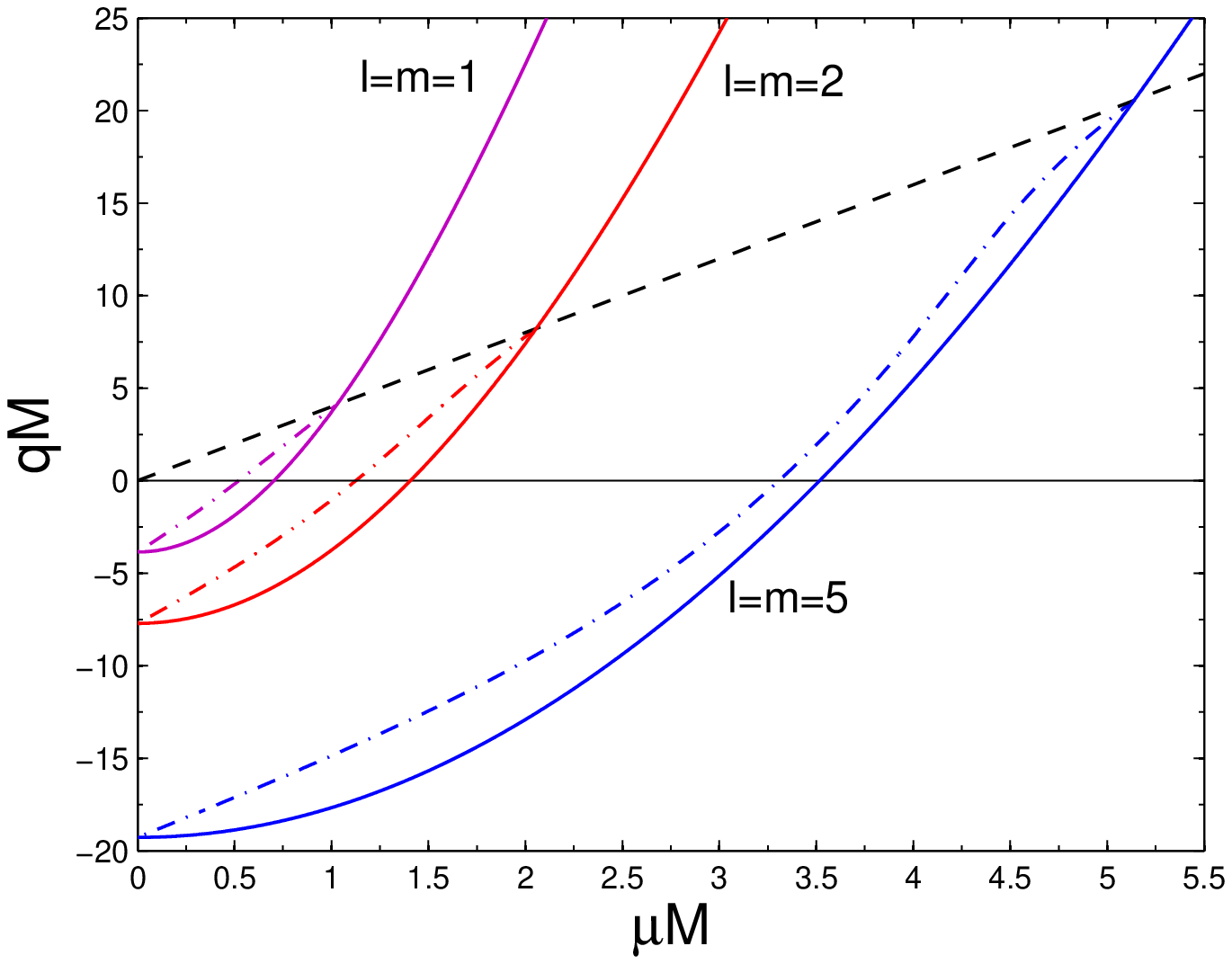}
 	}
 	\subfigure { \label{fig:1b}     
 		\includegraphics[width=0.45\textwidth]{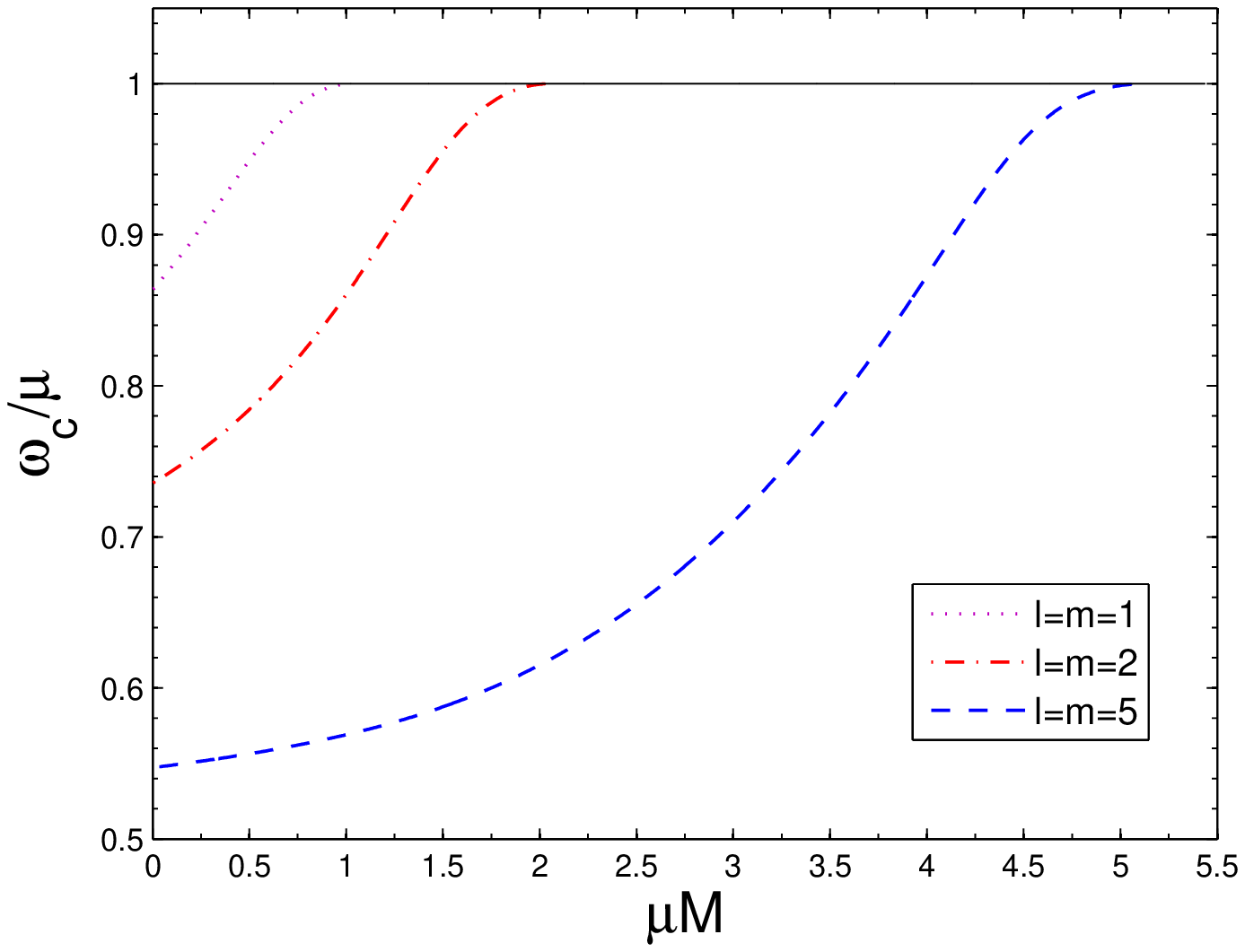} 
 	}
 	
 	\caption{Right panel: Existence lines (the dot-dashed lines) for nodeless ($n=0$)  scalar clouds with various quantum numbers in the $(\mu,q)$-parameter space of the scalar field. The dashed line denotes straight line (\ref{eq:fmu}) and the solid lines  denote hyperbolic curves  (\ref{eq:fomgC}). Left panel: The corresponding ratio  between critical frequency $\omega_c$ and mass $\mu$ of the scalar cloud as a function of $\mu$.  Here, the charge $Q$ and angular momentum $a$ of the black hole are fixed to be $0.25M$ and $0.968235M$, respectively.}
 	\label{fig:1}     
 \end{figure*}

\begin{figure*}    
 	\subfigure { \label{fig:2a}     
 		\includegraphics[width=0.45\textwidth]{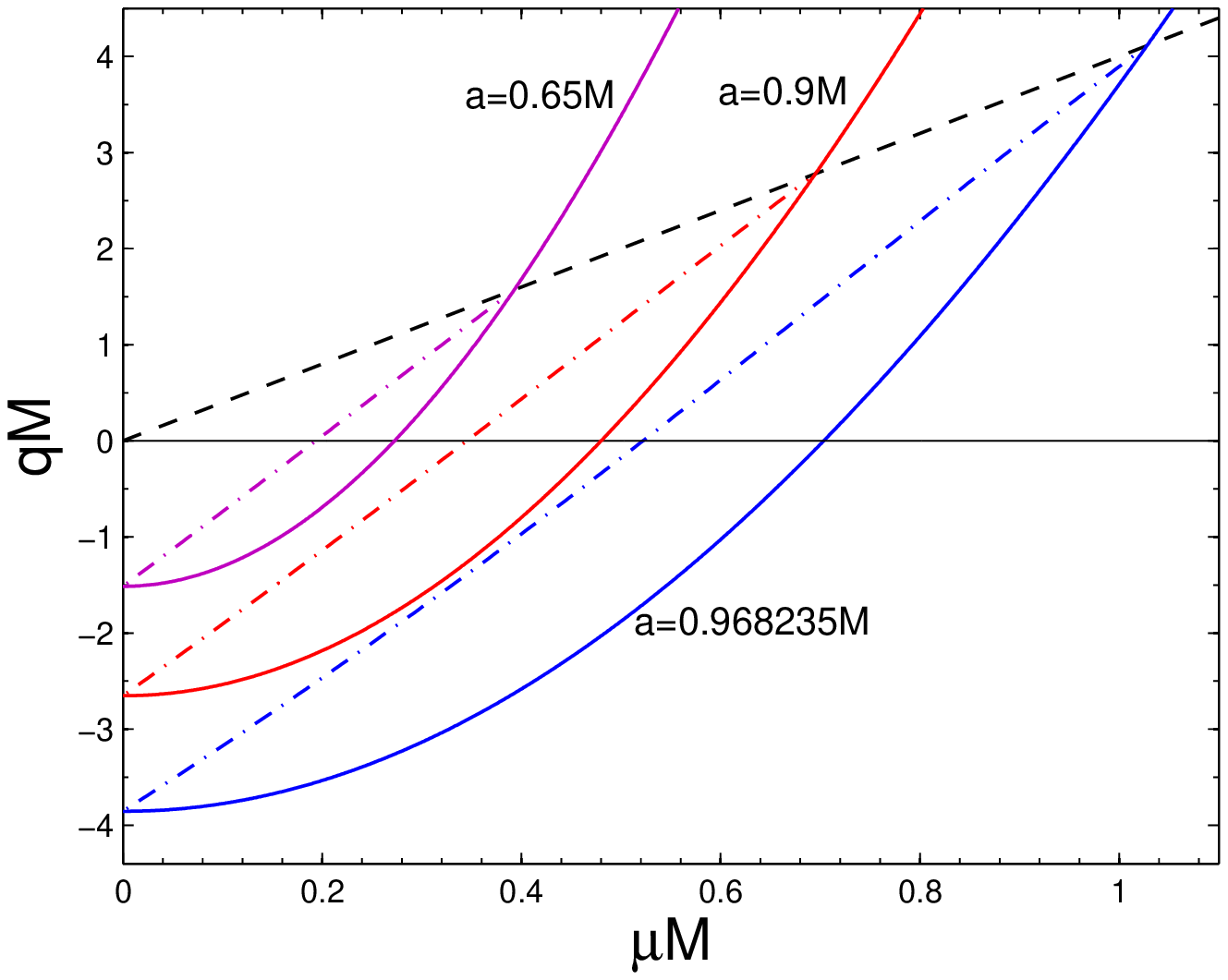} 
 	}
 	\subfigure { \label{fig:2b}     
 		\includegraphics[width=0.45\textwidth]{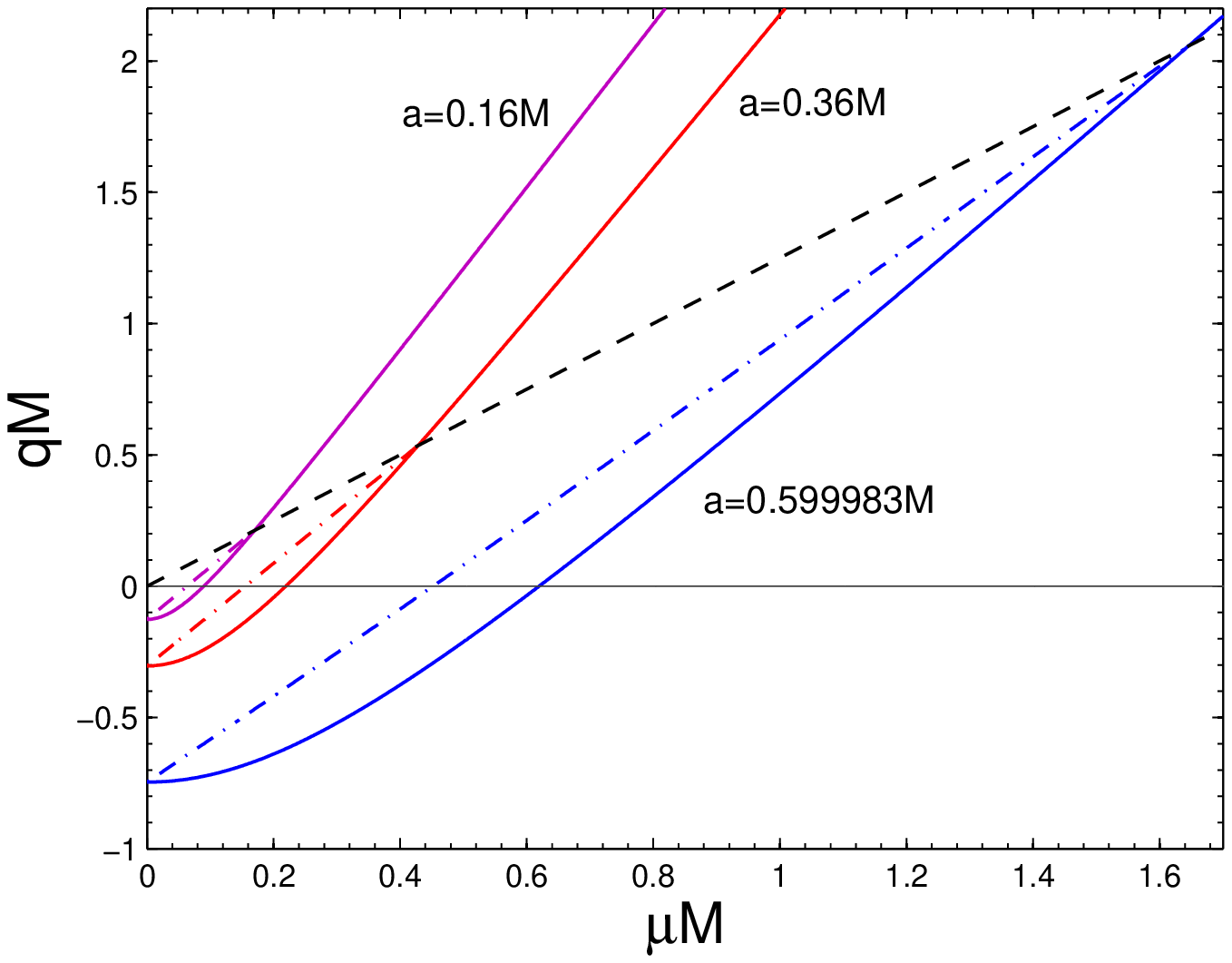}
 	}
 	\caption{Existence lines (the dot-dashed lines) for scalar clouds with given quantum numbers ($n=0,\,l=\,m\,=1$) in the $(\mu,q)$-parameter space of the scalar field for differ black hole parameters. The dashed lines and the solid lines denote straight line (\ref{eq:fmu}) and hyperbolic curves (\ref{eq:fomgC}), respectively. (Left panel: $Q=0.25M$; Right panel: $Q=0.8M$.) }
 	\label{fig:2}     
 \end{figure*}

 Our numerical results  are presented in Fig.\ref{fig:1} and Fig.\ref{fig:2}. In the two figures, the dot-dashed lines are the existence lines for scalar clouds. It is found that scalar clouds exist only within a limited region.  In particular,  for different sets of parameters,  the existence lines always join the two intersect points $P_1$ and $P_2$ (see Eqs.(\ref{P1}) and (\ref{P2})). Actually, in the process of computation, we find that the existence lines can not arrive at the two points, but can only approach infinitely to them. This is reasonable,  because  at the points $P_1$ and $P_2$,  $\mu$ equals $\omega_c$ (especially, $\mu=\omega_c=0$ at $P_2$), then from boundary condition (\ref{eqn:k}), if there exist scalar clouds with $\omega=\omega_c$ the clouds would stretch to spatial infinity.  From the right panel of Fig.\ref{fig:1}, we can see that as $\mu$ moves closer to its maximum value,  the ratio $\omega_c/\mu$ approaches to one from below as expected. 
 
Another interesting observation from Fig.\ref{fig:1} and Fig.\ref{fig:2} is that, for fixed black hole parameters, the cloud's charge $q$ grows monotonically as $\mu$ increases and when $\mu$ is relatively small, $q$ would take a negative value. Physically, there would be a simple explanation for this: when $\mu$ increases, it needs more Coulomb repulsion (a larger positive $q$) to maintain the equilibrium with the gravitational attraction; however, when $\mu$ is relatively small, the gravitational attraction is not enough to maintain stationary scalar configurations outside the Kerr-Newman black hole, a negative value of $q$ is thus needed.
 
 
 \subsection{Five partitions in the in the $(\mu,q)$-plane }
 
 Given three constraint lines and the existence line of scalar cloud discussed above, the $(\mu,q)$-plane can be divided into five partitions as illustrated in Fig.\ref{fig:paraplotorder}. Note that, here we are only interested in the region spanned  from $P_2$ to $P_1$ in the $(\mu,q)$-plane.  The physical properties of the system within various partitions  are summarized in Table \ref{table:1}. 
 It should be pointed out that, only when the three conditions (i.e, superradiance (\ref{superradiant condition}), potential well (\ref{eq:f1}) and exponential decaying (\ref{eq:omega_mu}))   are met,  it is possible to cause superradiant instability.  Therefore,  superradiant instability can only occur within partitions II and III.  

 \begin{table*}
 	\centering
 	\begin{tabular}{ |c|c|c|c|c|c| }
 		\hline
 		Partition & Case & Superradiance & Potential well & Exponential decaying & Superradiant instability\\ \hline
 		\multirow{4}{*}{I} & $\omega<\mu<f<\omega_c$ & Y & N & Y & N \\ 
 		& $\mu<\omega<f<\omega_c$ & Y & N & N & N \\ 
 		& $\mu<f<\omega<\omega_c$ & Y & Y & N & N \\ 
 		& $\mu<f<\omega_c<\omega$ & N & Y & N & N \\ 
 		\hline
 		\multirow{4}{*}{II} & $\omega<f<\mu<\omega_c$ & Y & N & Y & N \\ 
 		& $f<\omega<\mu<\omega_c$ & Y & Y & Y & Y \\ 
 		& $f<\mu<\omega<\omega_c$ & Y & Y & N & N \\ 
 		& $f<\mu<\omega_c<\omega$ & N & Y & N & N \\ 
 		\hline
 		\multirow{2}{*}{III} & $f<\omega<\omega_c<\mu$ & Y & Y & Y & Y \\ 
 		& $\omega<f<\omega_c<\mu$ & Y & N & Y & N \\ 
 		\hline
 		\multirow{2}{*}{IV} & $f<\omega_c<\omega<\mu$ & N & Y & Y & N \\ 
 		& $f<\omega_c<\mu<\omega$ & N & Y & N & N \\ 
 		\hline
 		\multirow{4}{*}{V} & $\omega<\omega_c<f<\mu$ & Y & N & Y & N \\ 
 		& $\omega_c<\omega<f<\mu$ & N & N & Y & N \\ 
 		& $\omega_c<f<\omega<\mu$ & N & Y & Y & N \\ 
 		& $\omega_c<f<\mu<\omega$ & N & Y & N & N \\ 
 		\hline
 	\end{tabular}
 	 	\caption{A summary on some physical properties of the scalar-field-black-hole system in various partitions. }
 	 	\label{table:1}
 \end{table*}
\section{Conclusion}
In this work,  we have investigated a massive charged scalar field linearly coupled to a charged rotating Kerr-Newman black hole.  For a given  black hole and a specific set of "quantum" numbers,  we have managed to divide the parameter space of the scalar field,  a plane spanned by its mass and charge,  into five partitions endowed with different physical properties  by three simple constraint lines and the existence line of scalar clouds.  The physical properties of the system in these partitions are presented. It is found that superradiant instability may be possibly caused only in two of the partitions.  In particular, we show that  both the mass and charge of the scalar clouds are bounded in a limited region. 

Our results can be used to rapidly judge the possible occurrence  of superradiant instability and  the existence of scalar clouds around a given black hole without solving a boundary value problem of differential equation. 
For instance, in the static limit $(a\rightarrow0)$,  Points $P_1$ and $P_2$ move closer together, and the space between tapers. If $a=0$, $P_1=P_2=(0, 0)$, which means the existence line will be degenerated and partitions II, III and IV will all have disappeared.  Thus, there is no superradiant instability in Reissner-Nordstr\"{o}m space-time \cite{PhysRevD.91.044047,Hod2012505,Hod20131489} and  no scalar clouds can be stably distributed around it.

Finally,  our research presented here is  limited to the scalar field minimally coupled to the gravitational and electromagnetic field of black hole.    
It would be of interest to generalize our study to the field with self-interaction or non-minimal couplings. Work in this direction will be reported in the future.

\begin{acknowledgments}
This work is supported in part by National Natural Science Foundation of China under Grant  No.~11275128, Shanghai Municipal  Commission of Science and technology under Grant No.~12ZR1421700 and the Program of Shanghai Normal University.
\end{acknowledgments}

\bibliography{cloud}
\end{document}